\documentclass[prd, preprint, aps, showpacs]{revtex4}

\usepackage{graphicx}
\usepackage{color}

\def\be{\begin{equation}}
\def\ee{\end{equation}}
\def\ba{\begin{eqnarray}}
\def\ea{\end{eqnarray}}
\def\beq{\begin{eqnarray}}
\def\eeq{\end{eqnarray}  }

\def\eref#1{Eq.~(\ref{#1})}

\def\rmd{{\rm d}}
\def\rmO{{\rm O}}
\def\ADM{{\rm ADM}}

\def\lp{\left(}
\def\rp{\right)}

\begin{document}
\title{Hyperbolicity and Constrained Evolution in Linearized Gravity}

\author{Richard A. Matzner}
\affiliation{Center for Relativity, University of Texas at Austin,
Austin, TX 78712-1081, USA}

\begin{abstract} Solving the 4-d Einstein equations as evolution in time 
requires solving equations of two types: the four elliptic initial data 
(constraint) equations, followed by the six second order evolution 
equations. Analytically the constraint equations remain solved under the 
action of the evolution, and one approach is to simply monitor them ({\it 
unconstrained} evolution). Since computational solution of differential 
equations introduces almost inevitable errors, it is clearly ``more correct" 
to introduce a scheme which actively maintains the constraints by solution 
({\it constrained} evolution). This has shown promise in computational 
settings, but the analysis of the resulting mixed elliptic hyperbolic method has 
not been completely carried out. We present such an analysis for one method 
of constrained evolution, applied to a simple vacuum system, linearized 
gravitational waves.

We begin with a study of the hyperbolicity of the unconstrained Einstein equations. 
(Because the study of hyperbolicity deals only with the highest derivative 
order in the equations, linearization loses no essential details.) We then 
give explicit analytical construction of the effect of initial data setting 
and constrained evolution for linearized gravitational waves. While this 
is clearly a toy model with regard to constrained evolution, certain 
interesting features are found which have 
relevance to the full nonlinear Einstein equations.

\end{abstract}

\pacs{ }

\maketitle

%
%
%
\section{Introduction}
\label{sec:intro}

Binary black hole systems are expected to be the strongest possible
astrophysical gravitational wave sources. In the final moments of
stellar mass black hole inspiral, the radiation will be detectable in
the current (LIGO-class) detectors. If the total binary mass is of the
order of $10M_{\odot}$, the moment of final plunge to coalescence will
emit a signal detectable by the current generation of detectors from
very distant (Gpc) sources. The merger of supermassive black holes in
the center of galaxies will be the very dominant signal in the
spaceborne LISA detector, and detectable out to large redshift.
Simulation of these mergers will play an important part in the
prediction, detection, and the analysis of their gravitational signals
in gravitational wave detectors. To do so requires a correct formalism which
does not generate spurious singularities during the attempted simulation.

Recent work at Texas \cite{gr-qc/0307055}, \cite{mattDiss}
has found that {\it constrained }3-d evolution can produce
substantially stabilized long-term single black hole 
simulations, stimulating interest in 
constrained evolution. However, it is true that to date there has 
been no analysis 
that addresses the behavior of constrained evolution for computational 
relativity. That is the purpose of this paper.
(Note that unlike the situation in electromagnetism, where a 
discrete simplectic calculus can be written which assures 
conservation of a discretized 
version of the E\&M constraints, there appears to be no 
equivalent formulation for 
general relativity that conserves discretized 
versions of the gravitational constraints\cite{Meier}.)

\section{$3+1$ Formulation of Einstein Equations}

We take a Cauchy formulation
(3+1) of the ADM type, after Arnowitt, Deser, and
Misner~\cite{ADM}. In such a method the 3-metric $g_{ij}$ and its momentum
$K_{ij}$ (the {\it extrinsic curvature}) are specified at one initial time on
a spacelike hypersurface, and evolved into the future.  The ADM metric is
\be
\rmd s^2 = -(\alpha^2 - \beta_i \beta^i)\,\rmd t^2 + 2\beta_i \, \rmd t
\,\rmd x^i
     + g_{ij}\, \rmd x^i\, \rmd x^j
\label{eq:admMetric}
\ee
where $\alpha$ is the lapse function and $\beta^i$ is the shift
3-vector;
functions that relate the coordinates on each hypersurface to each other.
\footnotemark

\footnotetext{Latin indices run $1,2,3$ and are lowered and raised by $ g_{ij}$
and its 3-d inverse $ g^{ij}$.}

The Einstein field equations contain both hyperbolic evolution equations,
and elliptic constraint equations.
The constraint equations for vacuum in the ADM decomposition are:
\beq
H = \frac{1}{2} [R - K_{ij}K^{ij} + K^2] &=& 0,
\label{eq:constraintH}
\eeq
\beq
H^i = \nabla_j \lp K^{ij} - g^{ij}K\rp  &=& 0.
\label{eq:constraintK}
\eeq
Here $R$ is the 3-d Ricci scalar constructed from the 3-metric, and $\nabla_j$ is the
3-d covariant derivative compatible with $ g_{ij}$.
Initial data must satisfy these constraint
equations; one may not freely specify all components of $g_{ij}$ and
$K_{ij}$.

The evolution equations from the
Einstein system are

\be
       \partial_t g_{ij} = -2\alpha K_{ij} +\nabla_j\beta_{i} + \nabla_i\beta_{j},
\label{eq:gdot}
\ee
and 

\be
      \partial_t  K_{ij} = -\nabla_i \nabla_j \alpha
+\alpha [R_{ij}-2K_{ia}K^{a}_{~j} + K K_{ij}]\\
+\beta^k K_{ij,k} +K_{kj}\beta^k_{,j}+K_{ik}\beta^k_{,j},
\label{eq:kdot}
\ee
where $R_{ij}$ is the 3-d Ricci tensor.

We refer to this form of the Einstein equations as of {\it ADM type},
referring to the fundamental development \cite{ADM}; this specific form is
called the $\dot g$ - $\dot K$ form. 
Here, \eref{eq:constraintH}--\eref{eq:constraintK}, the
constraint equations, are the vacuum Einstein equations ${}^4 G_{00} = 0$
and
${}^4 G_{0i} = 0$ respectively, for a 4-metric of 
form \eref{eq:admMetric}. \eref{eq:gdot}--\eref{eq:kdot}, the
evolution equations, are a first order form of the vacuum Einstein equations
${}^4 R_{ij} = 0$. The methods decribed here can be easily extended to 
linearized versions of some other formulations of Einstein's equations, 
and one of these is explicitly addressed below.

\section{Data Form}
\label{sec:DataForm}

In this paper we consider only linearized gravitational waves (but see Appendix A).

We write the spatial metric as 
$g_{ab}= \delta_{ab} +h_{ab}$. We work with zero shift vector: $\beta^i = 0$ 
(cf \cite{gr-qc/0402123}), but we consider a lapse $\alpha$ 
which may differ 
from the flat background value by a function of first order 
in the linearization: $\alpha = 1+ \alpha_1 +$\rm O(${h}_{ab}^2$). 
In particular we 
choose below a {\it densitized lapse}:
\begin{eqnarray}
\alpha &=& g^b   \nonumber \\
      &=& 1+ b (h_{xx}+h_{yy}+h_{zz}) +O({h}_{ab}^2), 
 \label{dens}
\end{eqnarray}
where the quantity $g$ is the determinant of the $3-$metric, and $b$ 
is a chosen constant. Generally, densitized lapse can involve 
multiplication by a fixed function of the coordinates, but we 
assume linearization from flat space, so this function here must be 
unity. 

For our linearized case, \eref{eq:gdot} reads 
\be
      \partial_t  h_{ij} = -2 K_{ij},
\label{eq:hdot}
\ee
and \eref{eq:kdot} is:
\begin{eqnarray}
      \partial_t  K_{ij} = &=& -\frac{1}{2}\alpha^2 \delta^{lm} \Bigl(\partial_l\partial_m
h_{ij} + \partial_i\partial_j h_{lm} -
\partial_i\partial_l h_{mj} - \partial_j\partial_l h_{mi}\Bigr) 
- \partial_i\partial_j \alpha.
\label{eq:linKdot}
\end{eqnarray}
\eref{eq:linKdot} 
can, with \eref{eq:hdot}, be 
read as a second-order in time equation for $h_{ij}$:
\begin{eqnarray}
\partial_t^2 h_{ij}&=& \alpha^2 \delta^{lm} \Bigl(\partial_l\partial_m
h_{ij} + \partial_i\partial_j h_{lm} -
\partial_i\partial_l h_{mj} - \partial_j\partial_l h_{mi}\Bigr) 
+ 2 \partial_i\partial_j \alpha,
 \label{adm3rd}
\end{eqnarray}

We further assume a $1-$dimensional system, so that all quantities are functions of 
$(x^1, t) = (x,t)$, where $x$ is a spatial variable. 
It is then profitable to write out the individual 
components explicitly. We shall 
find that the different components $xx$ (longitudinal-longitudinal: $LL$); 
$xy$ \& $xz$ (transverse-longitudinal: $TL$); $yy + zz$ (transverse trace); and   
$yy - zz$ \& $yz$ (transverse-traceless: $TT$) have, 
as expected, different behavior. This explicit formulation parallels, in a
less sophisticated but
more transparent manner, the decomposition in \cite{gr-qc/0402123}.

The second-order evolution equations, written explicitly in terms of these variables are:
\begin{eqnarray}
\partial_t^2 h_{xx}&=& {\partial_x}^2
(h_{yy}+h_{zz}) + 2
\partial_i\partial_j \alpha 
 \label{adm5th}
\end{eqnarray}
 \begin{eqnarray}
\partial_t^2 h_{xy}&=& 0
\label{adm3nd}
\end{eqnarray}
 \begin{eqnarray}
\partial_t^2 h_{xz}&=& 0
\label{adm3nda}
\end{eqnarray}
\begin{eqnarray}
\partial_t^2 (h_{yy}+h_{zz})&=& {\partial_x}^2
(h_{yy}+h_{zz})
 \label{adm6th}
\end{eqnarray}
\begin{eqnarray}
\partial_t^2 (h_{yy}-h_{zz})&=& {\partial_x}^2
(h_{yy}-h_{zz}) \label{adm7th}
\end{eqnarray}
\begin{eqnarray}
\partial_t^2 (h_{yz})&=& {\partial_x}^2
(h_{yz}).
\label{adm8th}
\end{eqnarray}

The constraint equations 
become:
\begin{eqnarray}
     H   &=& - {\partial}^2_x(h_{yy} + h_{zz}),
\label{HamLinExplicit}
\end{eqnarray}
\begin{eqnarray}
H_x &=& -2 {\partial}_x (K_{yy} + K_{zz}),
\label{HxLinExplicit}
\end{eqnarray}
\begin{eqnarray}
     H_y &=& 2 {\partial}_x K_{xy},
\label{HyLinExplicit}
\end{eqnarray}
\begin{eqnarray}
     H_z &=& 2 {\partial}_x K_{xz}.
\label{HzExplicit}
\end{eqnarray}

Note that since the spatial metric is $\delta_{ab}$, 
we drop the distinction between raised and lowered 
indices in these explicit Equations (\ref{adm5th})-(\ref{HzExplicit}).
We relate the constraint equations (\ref{HxLinExplicit})-
(\ref{HzExplicit}) to $ \partial_t  h_{ij}$ via \eref{eq:hdot}
above.

\section{Initial Data Setting and Constrained Evolution}

We first give the full, nonlinear data setting (constraint solving) procedure,
then specialize in the next section to our linearized plane wave case.

We adopt the conformal transverse-traceless method
of York and collaborators~\cite{YP}-\cite{Bowen+York} which consists of a
conformal decomposition with a scalar $\phi$ that adjusts the Hamiltonian 
constraint,
and a vector potential $w^i$ that adjusts the longitudinal components of the
extrinsic curvature.
The constraint equations are then solved for these new quantities such that
the complete solution fully satisfies the constraints.

For initial data setting we pose a trial metric and a trial trace-subtracted
extrinsic curvature 
taken as conformal trial functions  $\tilde{g}_{ij}$ and
$\tilde{A}^{ij}$.

The physical metric, $g_{ij}$, and the trace-free part of the extrinsic
curvature, $A_{ij}$, are related to the background fields through a
conformal
factor
\ba
g_{ij} &=& \phi^{4} \tilde{g}_{ij}, \label{confg1} \\
\label{confg}
A^{ij} &=& \phi^{-10} (\tilde{A}^{ij} + \tilde{(lw)}^{ij}),
\label{eq:conf_field}
\ea
where $\phi$ is the conformal factor, and $\tilde{(lw)}^{ij}$
will be used to cancel any possible longitudinal contribution in $\tilde{A}$.
$w^i$ is a vector potential, and
\ba
\tilde{(lw)}^{ij} \equiv \tilde{\nabla}^{i} w^{j} + \tilde{\nabla}^{j} w^{i}
        - \frac{2}{3} \tilde{g}^{ij} \tilde{\nabla_{k}} w^{k}.
\label{lw}
\ea
(Here $\tilde{\nabla_{k}}$ is the covariant derivative in the conformal 
background space.)
The trace $K$ is not corrected:
\be
K = \tilde K.
\label{tk}
\ee
Writing the full, nonlinear, Hamiltonian and momentum constraint 
equations in terms of
the quantities in 
Eqs.~(\ref{confg})--(\ref{tk}), we obtain four coupled
elliptic equations for the fields $\phi$ and $w^i$~\cite{YP}:
\ba
\tilde{\nabla}^2 \phi &=&  (1/8) \big( \tilde{R}\phi
        + \frac{2}{3} \tilde{K}^{2}\phi^{5} -   \nonumber \\
        & & \phi^{-7} (\tilde{A}{^{ij}} + (\tilde{lw})^{ij})
            (\tilde{A}_{ij} + (\tilde{lw})_{ij}) \big),  \label{phieq}
\ea
\ba
\tilde{\nabla}_{j}(\tilde{lw})^{ij} &=& \frac{2}{3} \tilde{g}^{ij} \phi^{6}
        \tilde{\nabla}_{j} K - \tilde{\nabla}_{j} \tilde{A}{^{ij}}.
\label{ell_eqs}
\ea

After the data have been set, the evolution begins. A well designed 
evolution scheme evolves to the next timestep. In an unconstrained 
evolution, this completes the time step. In our constrained scheme, 
the metric and extrinsic curvature so determined are taken 
as {\it intermediate} values {\it in the middle of a timestep} and 
thus as conformal trial functions  $\tilde{g}_{ij}$ and
$\tilde{A}^{ij}$.

The constraint Equations (\ref{phieq}),(\ref{ell_eqs}) are solved with these trial 
functions to complete each time update step.
The resulting solved $g_{ij}$ and $K_{ij}$ are taken as the data for the
next time update. Notice that these equations require no specific gauge
choice; the constraint solution is independent of the gauge functions 
$\alpha$ and $\beta^i$.
A similar approach also can be applied to other formulations which
generally have a larger number of constraints.

\section{Initial Data Setting for the Linearized System}

We assume a choice of $\tilde{h}_{ab}$ and $\tilde{K}_{ab}$.
We then solve the linearized version of the constraint 
Equations (\ref{phieq}), (\ref{ell_eqs}), obtaining $\phi$ 
and $w^i$. As in Section III we 
assume these variables are functions only of $x$ (not of time, 
because this is the intial data problem, which is solved at one instant of time). 
Then since the extrinsic curvature vanishes in the (flat) background, the 
linearized version of the Hamiltonian 
constraint (\ref{phieq}) is simply

\ba
{\partial}^2_x \phi &=&  \frac{1}{8}  \tilde{R},  \nonumber  \\
                     &=& -\frac{1}{8} {\partial}^2_x ({\tilde h}_{yy} + {\tilde h}_{zz} ).
\label{linear_phieq}
\ea
The linearized version of the momentum constraints (\ref{ell_eqs}), 
written explicitly for each component, are:
\ba
\frac{4}{3} w^{x,x}_{~~,x} &=& \frac{2}{3} 
       K_{,x} -  {\tilde A}^{xx}_{~~,x}.\\
w^{y,x}_{~~,x} &=&  -  {\tilde A}^{yx}_{~~,x}.\\
w^{z,x}_{~~,x} &=&  -  {\tilde A}^{zx}_{~~,x}.
\label{linear_w_eqs}
\ea

\subsection{Elliptic Equation Boundary Conditions}
\label{sec:boundary}

A solution of the elliptic constraint equations requires that boundary
data be specified. For this demonstration of the technique, analytic
integration in our 1-dimensional system
will lead to arbitrary integration constants. Specific
choices for these constants is equivalent to setting 
$\phi$ and $w^{i,x}$ to specific boundary values; this will be amplified in the 
Appendix, where we give a
brief discussion of the effect of such boundary condition choices. 
For instance, we will see that setting the integration 
constants to zero leads to simplest analytical results. However this is 
somewhat at variance with the process we use in full 3-dimensional simulations 
to handle black holes, where we choose conditions $\phi=1$ and $w^i=0$ at the
inner boundaries (the excision boundaries of the black holes); and mixed 
(Robin)\cite{Robin} conditions at the outer boundaries for black holes \cite{erin}.

\subsection{Constraint Solution for Linear Waves}
\label{sec:CSLW}

In the linearized regime the Equations (\ref{linear_phieq})-(\ref{linear_w_eqs}) 
have decoupled. 
Each of these equations is an ordinary differential equation, 
and is straightforwardly integrated. We obtain solutions:

\ba
 \phi_1 &=& -\frac{1}{8} ({\tilde h}_{yy} + {\tilde h}_{zz} )+Ax+B.
\label{eq:phiSoln}
\ea
\ba
\frac{4}{3} w^{x,x} &=& \frac{2}{3} 
       K -  \tilde A^{xx}+C_x,\\
\label{eq:wxxSoln}
w^{y,x} &=&  -  \tilde A^{yx}+C_y,\\
w^{z,x} &=&  -  \tilde A^{zx}+C_z,
\label{linear_ell_solutions}
\ea
where $A,B,C_i$ are integration constants.
Note that since $w^i$ is a potential, we need only its first derivatives.

>From \eref{confg}, the effect of the Hamiltonian constraint solution for $\phi$ is:
\ba
g_{ab} &=&  \delta_{ab} + h_{ab} \nonumber \\
       &=&  \exp{4 \phi}[\delta_{ab} + {\tilde h}_{ab}]\nonumber \\
       &=&  (1 + 4 \phi_1)[\delta_{ab} + {\tilde h}_{ab}] + O({h_{ab}}^2).
\ea

At the linearized level this conformal factor does not affect the off-diagonal components, 
so we obtain:
 
\ba
 h_{xx} &=& {\tilde h}_{xx} - \frac{1}{2} ({\tilde h}_{yy} + {\tilde h}_{zz} )
+4Ax+4B,\label{hsoln}\\
  h_{xy}      &=&  {\tilde h}_{xy},\\
  h_{xz}      &=&  {\tilde h}_{xz}, \\
  h_{yy} + h_{zz}    &=&  0 +8Ax+8B,\label{ttrace}\\
    h_{yy} - h_{zz}      &=&  {\tilde h}_{yy} - {\tilde h}_{zz},\\
        h_{yz}      &=&  {\tilde h}_{yz}.
\ea

Notice that the $TT$ (${yy} - {zz}$ \& ${yz}$)
components are unchanged, but {\it the transverse trace} 
(${yy} + {zz}$) {\it is set to zero} modulo integration constants, 
with a similar subtraction of the 
$LL$ $(xx)$ component, which, however, generally leaves $h_{xx}$ nonzero. 
With only this one 
variable $\phi$ it is not surprising that the $TL$ 
terms $( {zx}$ \& ${yx})$
are left unchanged.

The effect of solution of the momentum constraints is to modify 
the extrinsic curvature
by adding $\tilde{(\ell w)}^{ab}$. The components $w^y$ and $w^z$ set 
$A_{xy}$ and $A_{xz}$ to zero, respectively (modulo integration constants). 
$w^x$ modifies all the diagonal components
of $A_{ij}$: $A_{xx} = \tilde A_{xx}+\frac{4}{3} w^{x,x}$,
$A_{yy} =\tilde A_{yy}- \frac{2}{3} w^{x,x}$,
$A_{zz} =\tilde A_{zz}- \frac{2}{3} w^{x,x}$; note that $A_{ab}$ is traceless since 
$\tilde A_{ab}$ is. The
results: $A_{xx} = \frac{2}{3} K$, 
$A_{yy} = \tilde A_{yy} + \frac{1}{2}\tilde A_{xx}- \frac{1}{3} K $,
$A_{zz} = \tilde A_{zz} + \frac{1}{2}\tilde A_{xx}- \frac{1}{3} K. $
Written in terms of $K_{ab}$ 
rather than $A_{ab}$, and restoring the integration constants, we find:

\ba
 K_{xx} &=& K +C_x,\label{kxxCx} \\
 K_{xy} &=& 0 +C_y,\label{kxyCy} \\
 K_{xz} &=& 0 +C_z, \label{kxzCz} \\
 K_{yy} + K_{zz}    &=&  0 - C_x,\label{tKtrace}\\
 K_{yy} - K_{zz}    &=& {\tilde K}_{yy} - {\tilde K}_{zz},\\
 K_{yz}    &=& {\tilde K}_{yz}.
\label{ksoln}
\ea

Note, as assumed, that the trace, $K$, of the extrinsic curvature is not 
modified in the conformal constraint solution.

By inserting these results (\ref{hsoln})-(\ref{ksoln}) 
into Equations (\ref{HamLinExplicit})-(\ref{HzExplicit}), 
we can verify that the constraints are satisfied. 

\section{Free evolution and hyperbolicity}
\label{sec:FEH}

We now turn to the hyperbolicity of this system, assuming the data 
have been correctly set (the constraints are initially satisfied). 
Note that neither the $LL$ metric component $h_{xx}$ nor its 
initial time derivative is generically zero. 

Kreiss and Ortiz \cite{gr-qc/0106085} have shown that a second-order form,
and its equivalent first order form have the same hyperbolicity 
classification. In pure second order form (no first derivatives) the classification 
is  especially simple: If all the modes satisfy a wave equation 
with real nonzero propagation velocity, the system is strongly 
hyperbolic and well posed for Cauchy data. 
If any  of the modes has zero propagation velocity, the system 
is at best weakly hyperbolic and is
potentially ill posed. If any of the propagation velocities is 
imaginary the system is not 
hyperbolic, and is completely ill posed for Cauchy data.

If a unit lapse is chosen so that $\partial_x^2 \alpha =0$ then the $LL$ 
equation (\ref{adm5th}) contains no spatial derivatives 
of $h_{xx}$; the propagation velocity for $h_{xx}$ is zero. 
\eref{adm5th} then reflects at best weak hyperbolicity of the system; 
clearly it admits analytic linear growth in $h_{xx}$, which can 
plainly lead to late-time difficulties.
If, however, the lapse is densitized, with constant $b > 0$ (cf \eref{dens}), 
then \eref{adm5th} becomes wavelike for $h_{xx}$. If this were the only 
equation on the system this choice of lapse would have made the system strongly
hyperbolic.  

However, one must investigate the entire set of evolution equations. 
Note that the $TL$ components (\ref{adm3nd}), (\ref{adm3nda}) 
of the field equations have no spatial 
derivatives, again signaling weak hyperbolicity, and analytic linear-growth 
solutions. Though the initial data can 
analytically set the slope to zero (if we choose $C_y=C_z=0$), 
errors can arise in numerical 
evolution that triggers such linear growth.

Thus the $\dot g - \dot K$ version of these equations is weakly hyperbolic,
and questions can arise as to the well-posedness of the solutions, even with
densitized lapse. In ref\cite{gr-qc/0402123}, it is shown that a relatively simple 
extension that captures one of the main ideas of 
the BSSN\cite{BS}\cite{SN} approach
can cure the nonhyperbolicity of the $TL$ components. This involves 
considering the combination
\ba
 f_j &=& \delta^{kl} \partial_k h_{lj} - \frac{1}{2}\partial_k h. \label{fi}
\ea
In that case one must treat the first order form of the equations because 
those for $f_j$ are first order in time. The evolution equation for $f_i$ 
is determined by differentiation of its definition. However the result is 
still a weakly hyperbolic system. By subtracting a multiple of the momentum 
constraint $H_i$ to the right hand side of the equation for $\partial_t f_i$, a
strongly hyperbolic system can finally be achieved. 

We will take a 
somewhat different approach here, as we discuss in the next section.
The result just noted, that momentum constraint improves the 
hyperbolicity of the system, is of interest, because we will 
demonstrate momentum constraint {\it solution}, below. 

\section{constrained evolution}

Constrained evolution as we described in Section IV above modifies 
the evolution.
After each explicit step, the metric and extrinsic curvature (via \eref{eq:hdot})
are treated as conformal background quantities and are reset 
according to Eqs(\ref{hsoln})-(\ref{ksoln}).  Suppose in this 
Section that integration constants $A, B, C_i$ are all zero. 

The transverse trace, which might have 
grown an error from the discretization of the equation, is 
reset to zero; cf \eref{ttrace}. Also, the 
momentum of this component is reset to zero (\eref{ttrace}). 

The $h_{yy}+h_{zz}$ terms evolve to set the term $\tilde h_{yy}+\tilde h_{zz}$ 
to zero in \eref{hsoln}. In this case the only terms on the right side 
of the $\partial_t^2 h_{xx}$ equation are the second derivatives of 
the lapse, $\alpha$. In setting data the trace of the extrinsic curvature, 
$K_{initial}$, is one of the data functions to be set, and with constrained data 
(and zero integration constants) $K_{initial}$ gives $\partial_t h_{xx}$ 
via \eref{eq:hdot}.

With densitized lapse, the $h_{xx}$ equation (\ref{hsoln}) is effectively a wave 
equation for $h_{xx}$. Thus at least 
in this analytic discussion of the evolution, the 
linearized system appears completely controlled in a constrained, densitized-lapse
evolution. 

If the lapse is chosen constant, but constraint solution \eref{ttrace} is 
imposed, then \eref{adm5th} with \eref{kxxCx} shows the 
the evolution is {\it constant K} ($K=K_{xx}$ is constant in time, 
though $K_{,x}$ is not necessarily zero). While this causes no 
trouble in the 
linearized system, it will eventually violate the $|h_{xx}|<<1$ assumption. 
In the full 
nonlinear case this growth will eventually definitely cause problems. 
As we have noted,
the nondensitized case is definitely weakly-hyperbolic.

Finally, notice that in this case that constraint solution 
sets the $TL$ momenta $K_{xy}$ and $K_{xz}$ to zero and the 
TL momenta are 
reset to zero (Eqs.~(\ref{kxyCy})-(\ref{kxzCz})). Thus the 
remaining components contributing to weak hyperbolicity are 
controlled as a consequence of the momentum constraint. This 
is of interest because the method of \cite{gr-qc/0402123} required the 
addition of the momentum constraint to control the behavior 
of the $TL$ components in the evolution equations.
   
\section{Relation to Standard gauge Setting for Linearized waves}

In (``elementary") textbook 
formulations, it is common to say that a gauge can be set to put 
$h_{yy}+h_{zz}=0$ and $h_{xi}=0$ initially, and that this gauge is 
then maintained by the evolution equations. 
Those formulations are set in terms of $4-$dimensional gauge setting, 
rather than 
a $3+1$ split as we employ. We have seen that $h_{yy}+h_{zz} = 0$ is a 
requirement of the Hamiltonian constraint, modulo integration constants.
In our $3+1$ context, we then see that the $h_{xx}$ stays zero only if the 
initial 
slice is chosen $K=0$. 
This is of course a choice of the initial slice, so this is a 
$4-$dimensional 
gauge choice. 

This observation is relevant to the superficially very strange feature that, even 
just concentrating on the $h_{xx}$ equation, strong hyperbolicity requires 
densitized lapse. Mathematically this is perfectly well defined and consistent. 
To a physicist, however, it is extremely strange that we must warp our 
concept of $t=constant$ to achieve controllable solutions, and that this 
process actually accomplishes mathematically well behaved solutions. 
However, in the context of the observation that $K$ need not be initially 
set to zero, the situation becomes more comprehensible. The choice $K \ne 0 
$ is not consistent with the elementary textbook gauge setting just 
described. Instead by taking $K \ne 0 $ one has taken a peculiarly curved 
initial spatial slice; the specific choice of densitized lapse is required 
to correct it. While it may still seem strange that a wavelike (propagating) 
solution is the best way to accomplish it, this is what happens.

Further, another part of the assertion in elementary treatments is that the 
$TL$ components $h_{xy}$ and $h_{xz}$ be set to zero as a gauge choice, and 
will remain zero. The approach from hyperbolicity achieves this by 
introducing another equation with contribution on the right side 
proportional to the momentum constraint, which is ideally zero. The 
resulting system maintains evolution close to satisfaction of the gauge 
conditions. The constrained evolution approach guarantees exact solution of 
this gauge condition ($h_{xy}$ and $h_{xz}$ remain constant, and can be set 
to zero by a simple rotation around the $x-$axis). Because small errors can 
arise in any computational system, a combination of the two approaches seems 
the best approach to achieving long term computational relativity code 
stability.

\section*{Acknowledgments}

This work is supported in part by NSF grants PHY~0102204 and PHY~0354842.
Portions of this 
work were conducted at the Laboratory for 
High Energy Astrophysics, NASA/Goddard Space flight Center,
Greenbelt Maryland, with support from the University Space Research Association.  

\section*{Appendix A: Generality of Study of Hyperbolicity}

The analysis of the hyperbolicity in section \ref{sec:CSLW} is written in 
terms of linearized gravity as defined in section \ref{sec:DataForm}. 
However, based on arguments in \cite{gr-qc/0402123}, the results are 
completely general for ADM, whether posed in first or second order form. We 
now summarize arguments for several points from \cite{gr-qc/0402123}.

{\bf {\it $\beta =0$:}} For reasonable shift vectors (shift velocities less 
than $c$), ``the role of the shift vector is to displace the value of the 
eigenvalue... , and the lapse re-scales it. But a change of lapse... and 
shift can not change a real eigenvalue into an imaginary one. It can not 
affect the hyperbolicity of the system." See \cite{gr-qc/0402123}, section 
III A.

{\bf {\it Linearization:}} The hyperbolicity classification is based on the 
principal part of the symbol of the differential operator, which involves 
only the highest derivatives. Hence the terms shown in Eqs.(\ref{eq:hdot}), 
(\ref{eq:linKdot}), or equivalently \eref{adm3rd}, above define the 
hyperbolicity. This is true essentially because the Einstein equations are 
quasilinear, i.e. linear in the highest derivatives, and this is the 
reason that one can consider Fourier transforms of the linearized equations; 
consideration of hyperbolicity essentially looks at only the highest 
frequency behavior.

We close a defect in the statement just made by following an argument given 
by \cite {gr-qc/0402123}, ``since the norms $|~ |_\delta$ and $| ~|_h$ are 
equivalent and smoothly related, ... the properties of the 
eigenvalues and eigenvectors of the principal symbol are the same with 
either norm". Thus the undifferentiated metric (or inverse metric) coefficients 
in the Einstein equations can be equivalently  replaced by $\delta_{ij}$ or 
its inverse, leading to Eqs. (\ref{eq:hdot}), (\ref{eq:linKdot}) or the 
equivalent second order form \eref{adm3rd}.

{\bf {\it Restriction to $1-d$ Spatial Dependence:}} With the results just 
above, the background is isotropic, and nothing is lost in picking a 
specific direction for the wavevector $k_i$ associated with the local 
gradient.

Note that we do not claim generality for application of {\it constraint 
solution} beyond the linearized wave spacetime.

\section*{Appendix B: Other Boundary Conditions in Constraint Solution}

One of the advantages of writing explicit (if simple) solution of the 
linearized constraint equations is that one can explicitly investigate the 
effect of boundary conditions. For instance, in full $3-d$ constraint 
solution we impose the inner conditions $\phi=1$ and $w^i=0$, set 
at the excision surface of the black hole data. We take the outer conditions 
as in \cite{erin}; they  
are versions of the {\it Robin} \cite{Robin} condition. Thus for the black hole 
case the boundary condition for $\phi$ is taken as $\partial_r(r(\phi-1))=0$, and 
the outer boundary condition for the radial component of $w^i$, i.e. $w^r$, 
is taken as $\partial_r(rw^r)=0$. Define also the transverse components of $w^i$, 
$X^{jT} = w^i(\delta_i^j-\hat r_i \hat r^j)$ (where the $T$ means {\it 
transverse}). The outer boundary conditions on $X^{iT}$ are 
$\partial_r(r^2X^{iT})=0$.
We can model these boundary conditions in our $1-d$ system
by assuming a finite $x-$ domain, 
say $[0,x_m]$. (As in the $3-d$ situation, we will consider the outer
boundary $x_m$ becoming large, $x_m \rightarrow \infty$.)
We thus write  the boundary conditions in terms of $x-$ derivatives 
rather than $r-$ derivatives. Since we want to impose conditions on $w^i$ 
rather than its derivatives, the solution for $w^i_{~,x}$, 
Eqs. (\ref{eq:wxxSoln}) - (\ref{linear_ell_solutions})
can be integrated 
once again by writing the explicit ($0$ to $x$) integration of the source 
terms, and introducing new integration constants $D_i$. For instance

\ba
 w^y &=& -\int_{0}^{x} \tilde A^{yx}dx + C_y x +D_y
\ea

It is straightforward to verify that one can consistently set each of the 
variables $\phi_1,$ $w^i$ to zero at $x=0$, and satisfy the Robin conditions 
at $x=x_m$ as we now show.  (Notice that in our linearized problem, $\phi-1 = 
\phi_1$.)

We want  $w^y=0$ at $x=0$ and $\partial_x(x^2w^y)=0$ at $x=x_m$. Thus $D_y=0$ and

\ba
3C_y &=& \frac{2} {x_m}\int_{0}^{x_m} \tilde A^{yx}dx +\tilde A^{yx}(x_m).
\label{eq:cy}
\ea
A completely analogous treatment applies to $C_z$.

For $w^x$ we take conditions 
$w^x=0$ at $x=0$ (which gives
$D_x=0$), and $\partial_x(xw^x)=0$ at $x=x_m$, which gives 
\ba
2C_x&=&(\tilde A^{xx}- \frac{2}{3}K)(x_m) 
+\frac{1} {x_m} \int_{0}^{x_m}(\tilde A^{xx}- \frac{2}{3}K) dx.
\label{eq:cx}
\ea

The solution for $\phi_1$ directly involves two integration constants, 
$A,$ $B$, cf. \eref{eq:phiSoln}. We take inner boundary condition $\phi_1 = 0 $
at $x=0$, which sets $B = \frac{1}{8}(\tilde h_{yy} + \tilde h_{yy})|_0$. 
We then set $\partial_x (x\phi_1) =0$ at $x=x_m$, which gives

\ba 
16A&=& \partial_x(\tilde h_{yy} +\tilde h_{zz})(x_m) 
+ \frac{1}{x_m}(\tilde h_{yy} +\tilde h_{zz})|^{x_m}_0.
\label{aeq}
\ea

The ideal computational boundary condition has $x_m \rightarrow \infty$. 
With reasonable assumptions about the boundedness of the data at large 
$x_m$, Eqs.(\ref{eq:cy}) - (\ref{eq:cx}) lead to $C_i \rightarrow 0$
as $x_m \rightarrow \infty$. In \eref{aeq}, similarly $A \rightarrow 0$ as 
$x_m \rightarrow \infty$. 

Thus the $w^i $ have precisely the form assumed in the main text
with $C_i= 0$. For $\phi$ the situation is only slightly
mor complicated. We have

\ba 
\
\phi_1&=&- \frac{1}{8}(\tilde h_{yy} + \tilde h_{zz})|^x_0.
\label{phi1solneq}
\ea

>From \eref{ttrace} this gives the result for the metric components 
after the constraint solve:

\ba
  h_{yy} + h_{zz}    &=&  (\tilde h_{yy} +\tilde h_{zz})_0.\\
\ea

Such a constant value can be removed by a gauge (coordinate) transformation 
transverse to the $\hat x$ propagation direction. Hence we may take 
$(\tilde h_{yy} +\tilde h_{zz}) = 0$ in setting the data.

The result of the analysis in this Appendix is thus that the analysis 
in the main text accurately models the procedures used in our $3-d$
simulations. This has the important corollary that the boundary 
conditions in the $3-d$ case are completely internally consistent;
we have not set too restrictive conditions (which might have only isolated 
solutions, i.e. might define an eigenvalue problem).

It is not {\it necessary} to consider infinite domains $x_m \rightarrow \infty.$ 
To treat cosmologies for instance, it is more appropriate to consider 
a specific finite value of $x_m$. In this case the integration 
constants do not vanish. As an example, 
using \eref{ell_eqs} we find that the solution for $A^{xx} $ is
$A^{xx}= \frac{2}{3} K +C_x$.
Assuming that $K$ changes little from timestep to timestep, the constrained 
evolution reproduces this same value of $C_x$. Clearly $C_x$ can be 
consistently set to zero by setting $A^{xx}= \frac{2}{3} K$ exactly, 
regardless of the value of $x_m$

As above, we can interpret the integration constants from the Hamiltonian constraint  
as a kind of gauge choice in the 
initial data, which is re-imposed in the constrained evolution. For 
instance, the constants $A,$ $ B$, which arise from the Hamiltonian constraint, 
were set to zero in the body of the paper, implying that the transverse 
trace $h_{yy}+h_{zz}$ is set to zero. The terms
$8Ax+8B$ in \eref{hsoln} give constants to $h_{yy}+h_{zz}$, 
and the constant $C_x$ appearing in 
\eref{tKtrace} gives a time derivative to this transverse trace. Note that 
while $h_{yy}+h_{zz}$ satisfies a wave equation, \eref{adm6th}, none of the 
integration constant terms has a second spatial derivative. Hence they will 
evolve trivially:
\ba
h_{yy}+h_{zz}&=&8Ax+ 8B + 2C_xt.
\label{secular}
\ea
With this form, it can be seen that repeated application of the 
constraints, as in constrained evolution, consistently leads to the 
same value of the constant $A$. The form \eref{secular} identifies the 
terms associated with the integration constants as gauge terms (the gauge 
vector $\xi_x = 2A(y^2+z^2),$ $\xi_y = -4Axy,$ 
$\xi_z = -4Axz$ removes the time independent terms). The same method 
can be used to remove the time dependent term at any one time, but
this secular $C_xt$ term can eventually destroy the linearization 
assumption (unless $C_x$ is forced to zero as suggeted above). 
The secular term can be eliminated by an outgoing wave boundary condition, 
which forces $C_x=0$. 

Similar considerations apply to the terms $C_y $ and $C_z$ in 
Eqs. (\ref{kxyCy}), (\ref{kxzCz}). 

If the constraints are repeatedly applied, as in our 
form of constrained evolution, the value of $C_y$ is consistent. That is, 
from \eref{kxyCy}, $A_{xy} = K_{xy}$ has the constant value $C_y$. The evolution 
equation \eref{adm3nd} shows that $K_{xy}$ remains constant (ideally)
during an integration step. If $K_{xy}$ has the value $C_y$ as
integrated in \eref{kxyCy}, then \eref{eq:cy} consisitently returns 
the same value of $C_y$.
Clearly the behavior of $K_{xz}$ is similar. 
Again, such terms can cause secular growth, but such growth 
can be controlled by, for instance, putting an outgoing wave boundary 
condition on $h_{xy}$ and $h_{xz}$.

\end{document}